\documentclass[times,twocolumn,final]{elsarticle}

\usepackage{medima}
\usepackage{framed,multirow}

\usepackage{amssymb}
\usepackage{latexsym}

\usepackage{url}
\usepackage{xcolor}
\usepackage[hidelinks]{hyperref}
\definecolor{newcolor}{rgb}{.8,.349,.1}

\usepackage{mathtools}
\usepackage{amsmath}
\newcommand{\R}{\mathbb{R}}
\newcommand{\Z}{\mathbb{Z}}
\newcommand{\C}{\mathbb{C}}

\newcommand\Tstrut{\rule{0pt}{2.0ex}}         
\newcommand\Bstrut{\rule[-0.9ex]{0pt}{0pt}}   

\journal{Medical Image Analysis}

\begin{document}

\verso{B. Zhou \textit{et~al.}}

\begin{frontmatter}

\title{Dual-Domain Self-Supervised Learning for Accelerated Non-Cartesian MRI Reconstruction}

\author[1,2]{Bo \snm{Zhou}}
\ead{bo.zhou@yale.edu}
\author[1]{Jo \snm{Schlemper}}
\author[1,6]{Neel \snm{Dey}}
\author[1]{Seyed Sadegh \snm{Mohseni Salehi}}
\author[3]{Kevin \snm{Sheth}}
\author[2,4]{Chi \snm{Liu}}
\author[2,4,5]{James S. \snm{Duncan}}
\author[1]{Michal \snm{Sofka}\corref{cor1}}
\cortext[cor1]{Corresponding author.}
\ead{msofka@hyperfine.io}

\address[1]{Hyperfine Research, Guilford, CT, USA.}
\address[2]{Department of Biomedical Engineering, Yale University, New Haven, CT, USA.}
\address[3]{Department of Neurology, Yale School of Medicine, New Haven, CT, USA.}
\address[4]{Department of Radiology and Biomedical Imaging, Yale School of Medicine, New
Haven, CT, USA}
\address[5]{Department of Electrical Engineering, Yale University, New Haven, CT, USA.}
\address[6]{Computer Science and Engineering, New York University, New York, NY, USA.}






\begin{abstract}
While enabling accelerated acquisition and improved reconstruction accuracy, current deep MRI reconstruction networks are typically supervised, require fully sampled data, and are limited to Cartesian sampling patterns. These factors limit their practical adoption as fully-sampled MRI is prohibitively time-consuming to acquire clinically. Further, non-Cartesian sampling patterns are particularly desirable as they are more amenable to acceleration and show improved motion robustness. To this end, we present a fully self-supervised approach for accelerated non-Cartesian MRI reconstruction which leverages self-supervision in both k-space and image domains. In training, the undersampled data are split into disjoint k-space domain partitions. For the k-space self-supervision, we train a network to reconstruct the input undersampled data from both the disjoint partitions and from itself. For the image-level self-supervision, we enforce appearance consistency obtained from the original undersampled data and the two partitions. Experimental results on our simulated multi-coil non-Cartesian MRI dataset demonstrate that DDSS can generate high-quality reconstruction that approaches the accuracy of the fully supervised reconstruction, outperforming previous baseline methods. Finally, DDSS is shown to scale to highly challenging real-world clinical MRI reconstruction acquired on a portable low-field (0.064 T) MRI scanner with no data available for supervised training while demonstrating improved image quality as compared to traditional reconstruction, as determined by a radiologist study.
\end{abstract}


\end{frontmatter}


\section{Introduction}

Magnetic resonance imaging (MRI) is a common medical imaging modality for disease diagnosis \citep{vlaardingerbroek2013magnetic}. However, MRI is inherently challenging due to its slow acquisition arising from physical and physiological constraints, with real-world scan times ranging from 15 mins to over an hour depending on the protocol and diagnostic use-case. Prolonged MR imaging sessions are impractical as they lead to increased patient discomfort and increased accumulation of motion artifacts and system imperfections in the image. Consequently, there is significant interest in accelerating MRI acquisition while maintaining high image fidelity.

MRI is typically acquired by sequentially sampling the frequency-domain (or \textit{k-space}) in a pre-defined pattern. For Cartesian MRI, a k-space grid is regularly sampled and an inverse Fourier transform may be directly applied to reconstruct the image (assuming that the Nyquist sampling rate is met). However, accelerated MRI generally uses non-Cartesian sampling patterns, such as spiral \citep{delattre2010spiral}, radial \citep{knoll2011second}, variable density \citep{knoll2011adapted}, and optimized sampling patterns \citep{lazarus2019sparkling}. Advantages of non-Cartesian sampling patterns include enabling more efficient coverage of k-space \citep{wright2014non} and enhanced patient motion robustness \citep{forbes2001propeller,pipe1999motion}. Further, recent accelerated MRI reconstruction studies have also shown non-Cartesian sampling to be better suited towards compressed sensing (CS) \citep{lustig2008compressed} and potentially deep learning (DL) based reconstructions as aliasing artifacts from non-Cartesian sampling shows higher noise-like incoherence than Cartesian sampling. In this work, we focus on the reconstruction of non-Cartesian sampled data. 

To accelerate MRI acquisition, various efforts have been made for reconstructing high-quality images with undersampled k-space data. The previous methods can be summarized into two categories: CS-based reconstruction \citep{lustig2008compressed} and DL-based reconstruction \citep{chandra2021deep}. CS-based reconstruction methods typically use sparse coefficients in transform-domains (e.g. wavelets \citep{qu2010combined,zhang2015exponential}) with application-specific regularizers (e.g. total variation) to solve the ill-posed inverse problem in an iterative fashion \citep{liang2009accelerating}. However, iterative sparse optimization approaches are prone to reconstructing over-smoothed anatomical structure and may yield undesirable image artifacts, especially when the acceleration factor is high (acceleration factor $>3$) \citep{ravishankar2010mr}. Moreover, iterative optimization-based reconstruction is time-consuming and requires careful parameter tuning across different scanners and protocols and may even require subject-specific tuning.

With recent advances in computer vision and the availability of large-scale MRI datasets \citep{zbontar2018fastmri,bien2018deep}, DL-based reconstruction methods have demonstrated significant improvements over CS-based methods. However, most DL-based reconstruction methods are limited to Cartesian sampling patterns and are supervised, thus requiring paired fully-sampled acquisitions for ground-truth \citep{wang2016accelerating,sun2016deep,schlemper2017deep,qin2018convolutional,zhu2018image,hammernik2018learning,quan2018compressed,lonning2019recurrent,han2019k,li2019segan,zhang2019reducing,zhou2020dudornet,liu2021deep,feng2021dual,dar2020transfer}. However, these requirements are impractical as several real-world MRI use-cases may not have the time/resources to fully sample k-space for supervised training or may prefer non-Cartesian sampling for its motion robustness advantages, among others. For example, full k-space sampling cannot be done for real-time cardiac MRI \citep{coelho2013mr} and functional brain MRI \citep{bagarinao2006real} where data acquisition time frames are tightly restricted.

Currently, there exist a handful of DL-based reconstruction methods to address non-Cartesian sampling while still requiring paired fully-sampled data for supervision \citep{aggarwal2018modl,schlemper2019nonuniform,ramzi2022nc,ramzi2021density}. Further, to obviate the need for supervised training and fully-sampled data, \textit{self-supervised} MRI reconstruction methods have been recently proposed where reconstruction networks can be trained without fully-sampled data \citep{wang2020neural,yaman2020self,yaman2021zero,acar2021self,hu2021self,martin2021physics,cole2020unsupervised}. However, these methods are still limited to Cartesian sampling patterns and do not address how to perform self-supervised learning for accelerated non-Cartesian MRI reconstruction.

\begin{figure*}[htb!]
\centering
\includegraphics[width=0.86\textwidth]{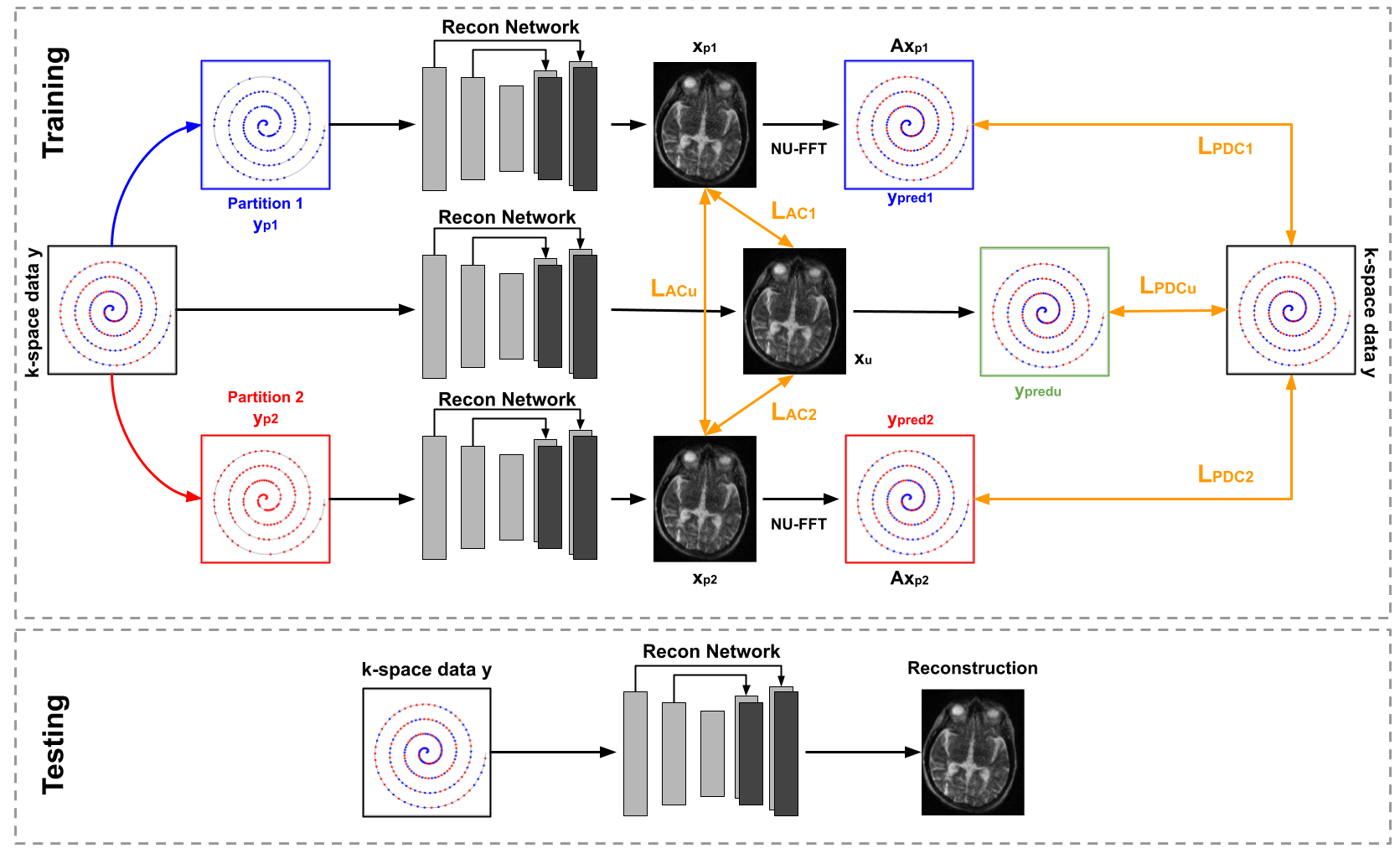}
\caption{\textbf{Dual-Domain Self-Supervised (DDSS) learning}. In training, undersampled k-space data $y$ is randomly partitioned into disjoint sets $y_{p_{1}}$ and $y_{p_{2}}$. 
$y$, $y_{p_{1}}$, and $y_{p_{2}}$ are then fed into the network to produce image-domain reconstructions ($x_{p_{1}}$, $x_{p_{2}}$, and $x_u$, respectively) and k-space reconstructions ($y_{pred_{1}}$, $y_{pred_{2}}$, and $y_{pred_{u}}$, respectively).
These outputs are trained under dual-domain losses $\mathcal{L}_{PDC_{1, 2, u}}$ (for k-space consistency) and $\mathcal{L}_{AC_{1, 2, u}}$ (for appearance consistency). Once trained, the trained network can directly reconstruct the image from $y$.}
\label{fig:pipeline}
\end{figure*}

In this work, we present Dual-Domain Self-Supervised (DDSS) reconstruction, a self-supervised learning method for accelerated non-Cartesian MRI reconstruction. DDSS is trained only on non-Cartesian undersampled data by first randomly partitioning the input undersampled k-space acquisitions into two disjoint sets with no overlap in k-space coordinates. The network is then trained under a combination objective function (Fig. \ref{fig:pipeline}) exploiting: (1) k-space self-similarity, where the network is trained to reconstruct the input undersampled k-space data from both the input and from the two disjoint partitions and; (2) reconstruction self-similarity, where the network is trained to enforce appearance consistency between the reconstructions obtained from the original undersampled data and the two partitions.

Experimentally, as existing large-scale MRI datasets use Cartesian acquisitions \citep{zbontar2018fastmri,bien2018deep}, we first simulate a multi-coil accelerated non-Cartesian dataset with spatial and phase degradations from the publicly available Human Connectome Project dataset~\citep{van2013wu} and use the simulated complex-valued images as ground truth for evaluation.
We find that DDSS generates accurate reconstructions that approach the fidelity of full supervision and outperform a series of baseline methods and ablations. We then demonstrate the successful application of DDSS in reconstructing challenging real-world MRI data from a clinically deployed portable low-field (0.064 T) MRI scanner \citep{Mazurek2021}, which acquires k-space data with a non-Cartesian variable density sampling pattern with no fully sampled references available. In this case, a survery of multiple radiologists indicates DDSS improvements in terms of image quality as compared to the reconstruction algorithm deployed in the system. Our major contributions include:
\begin{itemize}
  \item A self-supervised learning method that enables training deep reconstruction networks for non-Cartesian MRI without access to fully sampled data by exploiting dual-domain self-supervision in both k-space and image domains via novel loss functions and training strategies.
  \item Successful deployment (as measured by fidelity metrics, radiologist opinions, and improvements on previous work) on a simulated non-Cartesian accelerated MRI dataset and on real-world low-field 0.064T MRI scanners where fully sampled data is practically infeasible.
\end{itemize}

\section{Related Work}
\noindent\textbf{Fully-Supervised MRI Reconstruction.} 
With the recent advances in medical computer vision and the availability of large-scale paired Cartesian MRI datasets such as fastMRI \citep{zbontar2018fastmri} and MRNet \citep{bien2018deep}, significant effort has been made towards the development of fully supervised deep networks for accelerated Cartesian MRI reconstruction. \citet{wang2016accelerating} proposed the first DL-based MR reconstruction solution for training a three-layer convolutional network to map accelerated zero-filled reconstructions to fully sampled reconstructions. \citet{sun2016deep} developed ADMM-Net, which included an iterative ADMM-style optimization procedure into a deep reconstruction network. \citet{schlemper2017deep} proposed to use a deep cascade network with data consistency layers to approximate the closed-form solution of the iterative reconstruction. \citet{qin2018convolutional} and \citet{lonning2019recurrent} further advanced this network design with recurrent components to improve the reconstruction quality. \citet{hammernik2018learning} proposed to unroll the iterative optimization steps into a variational network. \citet{li2019segan} and \citet{quan2018compressed} suggested adversarial learning for improving anatomical details and structure in the reconstruction. Recently, \citet{feng2021dual} aggregated spatial-frequency context with a dual-octave convolution network to further improve reconstruction fidelity. 

\begin{figure*}[!ht]
\centering
\includegraphics[width=0.88\textwidth]{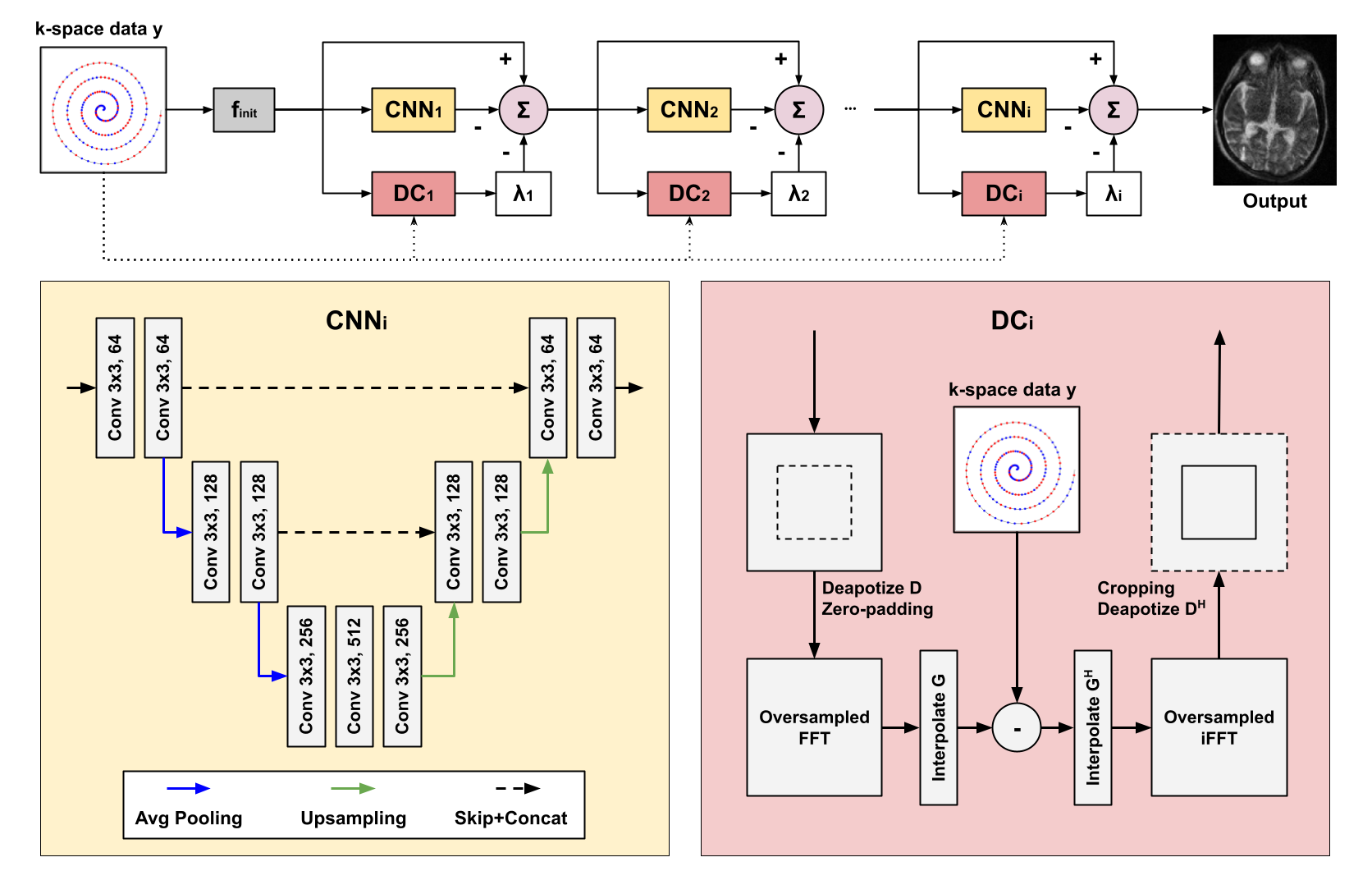}
\caption{\textbf{Variational reconstruction network architecture}. The variational network is a unrolled network to approximate gradient descent on equation \ref{eq:optimize}. The backbone network is shown in the yellow block and the data consistency (DC) operation is shown in the red block.}
\label{fig:network}
\end{figure*}

In addition to image domain-based reconstruction, \citet{han2019k} proposed to restore missing k-space measurements with deep models allowing the direct application of the inverse Fourier transform for image reconstruction. Similarly, \citet{zhu2018image} proposed to directly learn a mapping between undersampled k-space measurements to  the image domain. Combining image and k-space domains, \citet{zhou2020dudornet} showed that dual-domain recurrent learning can further improve reconstruction as compared to learning reconstruction in a single-domain only. \citet{liu2021deep} and \citet{zhou2020dudornet} also demonstrated that using multi-modal MRI input into the reconstruction network improves anatomical accuracy in the reconstructions. To directly address the issue of having insufficient data in the target domain of MRI reconstruction, transfer-learning methods have also been investigated to perform training in a data-abundant domain (for example, natural images), and then to fine-tune reconstruction models with few samples in the target domain \citep{dar2020transfer}.

However, the aforementioned works focus on MRI reconstruction with Cartesian sampling patterns, with only a limited number of methods developed for non-Cartesian sampling. For example, \citet{aggarwal2018modl} proposed a variational network with a conjugate gradient-based data consistency block that is suitable for non-Cartesian MRI reconstruction. Similarly, \citet{schlemper2019nonuniform} developed a gradient descent-based variational network for non-Cartesian MRI reconstruction. More recently, \citet{ramzi2022nc} also designed a non-Cartesian density compensated unrolled network, called NC-PDNet, for high-quality non-Cartesian accelerated MRI reconstruction. However, again, these methods use supervised training from large-scale paired non-Cartesian MRI which may be hard or infeasible to obtain in the real world or on new scanners. We address these challenges by taking a self-supervised reconstruction approach using only undersampled non-Cartesian data without paired ground truth. \\

\noindent\textbf{Self-Supervised MRI Reconstruction.} To alleviate the dependence on pairs of undersampled and fully sampled k-space measurements for MRI reconstruction training, self-supervised MRI reconstruction methods have been recently explored. For example, \citet{wang2020neural} proposed HQS-Net which decouples the minimization of the data consistency term and regularization term in \cite{schlemper2017deep} based on a neural network, such that the deep network training relies only on undersampled measurements. \citet{cole2020unsupervised} developed a self-supervised GAN-based reconstruction method, called AmbientGAN in which the discriminator is trained to differentiate between real undersampled k-space measurements from k-space measurements of a synthesized image, such that the generator learns to generate fully sampled measurements. \citet{yaman2020self,yaman2021zero} proposed a physically-guided self-supervised learning method, called SSDU, that partitions the undersampled k-space measurements into two disjoint sets and trains the reconstruction network \citep{schlemper2017deep} by predicting one k-space partition using the other. This method has also been successfully applied to dynamic MRI reconstruction~\citep{acar2021self}. Extending SSDU, \citet{hu2021self} proposed a parallel self-supervised framework towards improved accelerated Cartesian MRI reconstruction. On the other hand, instead of partitioning the undersampled k-space measurements into two disjoints as done in SSDU, \citet{martin2021physics} proposed to train the reconstruction network by predicting the undersampled k-space measurement with the same undersampled measurement as network input. Lastly, \citet{korkmaz2022unsupervised} presented an adversarial vision transformer for unsupervised patient-specific MRI reconstruction.

However, these methods mainly focus only on Cartesian MRI reconstruction and heavily rely on k-space data and thus do not take full advantages of the self-supervised learning in the image domain. In this work, we take advantage of \textit{both} k-space and image domain self-supervision and develop a dual-domain self-supervised learning method for accelerated \textit{non-Cartesian} MRI reconstruction. 

\section{Methods}
\subsection{Problem Formulation}
Let $x \in \C^N$ be a complex-valued 2D image to be reconstructed, where $x$ is a vector with size of $N=N_x N_y$ and $N_x$ and $N_y$ are the height and width of the image. Given a undersampled k-space measurement $y \in \C^M (M << N)$, our goal is to reconstruct $x$ from $y$ by solving the unconstrained optimization problem,
\begin{equation} \label{eq:optimize}
    \underset{x}{\arg\min} \quad \frac{\lambda}{2} || Ax - y ||^2_2 + R(x) ,
\end{equation}
where $A$ is a non-uniform Fourier sampling operator, and $R$ is a regularization term on reconstruction. If data is acquired under Cartesian sampling patterns, then $A = MF$, where $M$ is a sampling mask with the same size as $A$ and $F$ is the discrete Fourier transform. If data is acquired under a non-Cartesian sampling pattern, the k-space measurement locations will no longer located on a uniform k-space grid and thus a generalized definition of $A$ can be given by the non-uniform discrete Fourier transform:
\begin{equation}
    y((k_x, k_y)) = \sum_{w=0}^{N_x} \sum_{h=0}^{N_y} x_{wh} e^{2 \pi i (\frac{w}{N_x}k_x + \frac{h}{N_y}k_y)} ,
    \label{eq:nufft}
\end{equation}
where $(k_x, k_y) \in \R^2$, in contrast to $(k_x, k_y) \in \Z^2$ under Cartesian sampling. With the non-uniform Fast Fourier Transform (NUFFT) \citep{fessler2003nonuniform,greengard2004accelerating}, equation \ref{eq:nufft} can be approximated by:
\begin{equation}
    A = G F_s D ,
\end{equation}
where $G$ is a gridding interpolation kernel, $F_s$ is the fast Fourier Transform (FFT) with an oversampling factor of $s$, and $D$ is the de-apodization weights. The inversion of $A$ under fully-sampled cases can be approximated by gridding reconstruction:
\begin{equation} \label{eq:gridding}
    x = A^H W y
\end{equation}
where $W$ is a diagonal matrix for the density compensation of irregularly spaced measurements. However, when undersampled, this inversion is ill-posed, thus requiring one to solve equation \ref{eq:optimize}.

\begin{figure*}[htb!]
\centering
\includegraphics[width=0.999\textwidth]{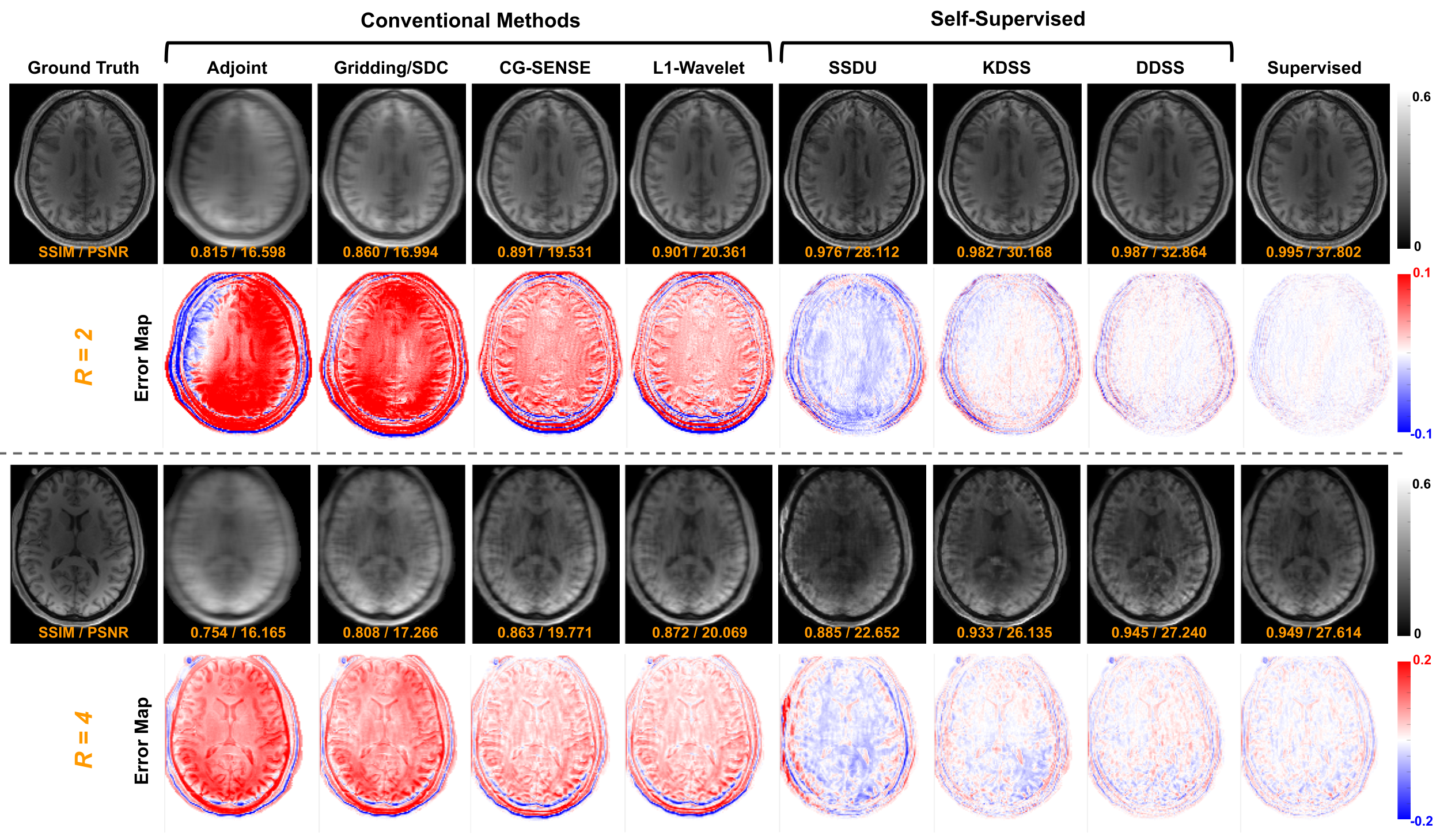}
\caption{T1w MRI reconstruction results on a simulated non-Cartesian degraded and undersampled dataset across all benchmarked methods. Both R=2 (top) and R=4 (bottom) acceleration results are shown alongside their respective error maps. Closer to a blank color (error map) indicates better performance.}
\label{fig:compare_simulation_T1_methods}
\end{figure*}

\begin{figure*}[htb!]
\centering
\includegraphics[width=0.999\textwidth]{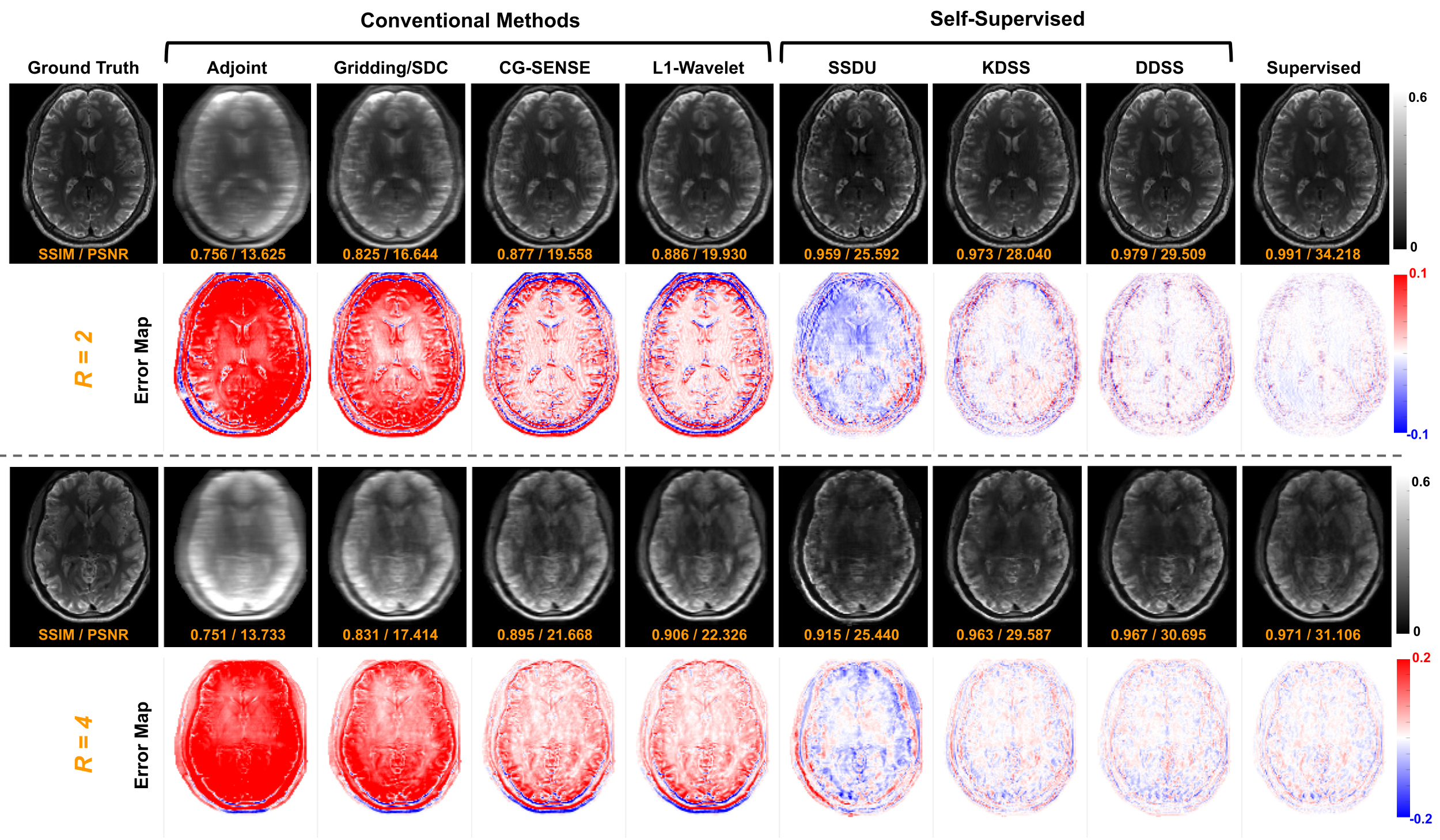}
\caption{T2w MRI reconstruction results on a simulated non-Cartesian degraded and undersampled dataset across all benchmarked methods. Both R=2 (top) and R=4 (bottom) acceleration results are shown alongside their respective error maps. Closer to a blank color (error map) indicates better performance.}
\label{fig:compare_simulation_T2_methods}
\end{figure*}

\subsection{Non-Cartesian Reconstruction Network}
In this work, we use a variational network \citep{schlemper2019nonuniform,hammernik2018learning} to approximate the solution to the optimization in equation \ref{eq:optimize}. The network structure and detailed optimization steps are illustrated in Fig. \ref{fig:network}. Using a gradient descent algorithm, a locally optimal solution to equation \ref{eq:optimize} can be iteratively computed as,
\begin{equation}
    x_i = x_{i-1} - \alpha_i \nabla_x f(x)|_{x=x_{i-1}},
\end{equation}
with an initial solution of:
\begin{equation}
    x_0 = f_{init}(A, y),
\end{equation}
where $f_{init}$ is an initialization function that is set as $f_{init}(A,y) = A^H y$. $\alpha_i$ is the gradient descent step size, and $\nabla f$ is the gradient of the objective function,
\begin{equation}
    \nabla_x f(x) = \lambda A^H (Ax - y) + \nabla_x R(x).
\end{equation}
We unroll the sequential update steps and formulate it as a deep learning-based feed-forward model, in which the regularization gradient term $\nabla_x R(x)$ is approximated by neural network. Here, we used a 3-level UNet \citep{ronneberger2015u} to approximate it. Thus, we have an end-to-end trainable variational network with $N_{iter}$ blocks:
\begin{equation}
    x_i = x_{i-1} - \lambda_i A^H (Ax_{i-1} - y) + f_{cnn} (x_{i-1}|\theta_i)
\end{equation}
where $\theta$ and $\lambda$ are learnable parameters. The second term is the data consistency (DC) term, and the third term is the CNN term.

\begin{table*} [htb!]
\small
\centering
\caption{Quantitative comparisons of T1w and T2w reconstructions with ground-truth reconstructions available under two different non-Cartesian acceleration settings using SSIM, PSNR, and NMSE (NMSE are scaled up by $10^3$). Best and second best results are \underline{underlined} and marked in \textbf{bold}, respectively. "$\dagger$" indicates that the difference between DDSS and all conventional and self-supervised baseline methods are significant at $p < 0.00833$ (Bonferroni multiple comparisons-adjusted alpha level) based on the non-parametric Wilcoxon signed rank test.}
\label{tab:comp_methods}
    \begin{tabular}{l|c|c|c||c|c|c}
        \hline
        \textbf{Evaluation}      & \multicolumn{3}{c}{\textbf{Setting 1 (R=2)}}              &  \multicolumn{3}{c}{\textbf{Setting 2 (R=4)}}                        \Tstrut\Bstrut\\
        \cline{2-7}
        \textbf{T1w}             & SSIM             & PSNR              & NMSE               & SSIM             & PSNR              & NMSE                      \Tstrut\Bstrut\\
        \hline
        Adjoint                  & $0.810 \pm 0.057$  & $15.881 \pm 4.150$  & $0.848 \pm 0.663$    & $0.762 \pm 0.071$  & $14.584 \pm 4.192$  & $1.174 \pm 0.934$         \Tstrut\Bstrut\\
        Gridding/SDC             & $0.886 \pm 0.032$  & $19.941 \pm 3.767$  & $0.295 \pm 0.145$    & $0.818 \pm 0.057$  & $16.859 \pm 4.097$  & $0.675 \pm 0.555$         \Tstrut\Bstrut\\
        CG-SENSE                 & $0.921 \pm 0.021$  & $22.641 \pm 3.230$  & $0.151 \pm 0.049$    & $0.880 \pm 0.034$  & $19.997 \pm 3.450$  & $0.291 \pm 0.148$         \Tstrut\Bstrut\\
        L1-Wavelet               & $0.927 \pm 0.018$  & $23.106 \pm 3.091$  & $0.134 \pm 0.037$    & $0.893 \pm 0.028$  & $20.787 \pm 3.280$  & $0.238 \pm 0.109$         \Tstrut\Bstrut\\
        \hline
        SSDU                     & $0.960 \pm 0.021$  & $26.301 \pm 1.253$  & $0.073 \pm 0.041$    & $0.898 \pm 0.033$  & $23.370 \pm 2.911$  & $0.123 \pm 0.019$         \Tstrut\Bstrut\\
        KDSS                     & $0.981 \pm 0.004$  & $30.454 \pm 2.447$  & $0.024 \pm 0.005$    & $0.941 \pm 0.022$  & $26.303 \pm 3.169$  & $0.063 \pm 0.011$         \Tstrut\Bstrut\\
        DDSS (Ours)              & $\mathbf{0.988 \pm 0.006}$$^\dagger$  & $\mathbf{33.678 \pm 3.926}$$^\dagger$  & $\mathbf{0.012 \pm 0.005}$$^\dagger$    & $\mathbf{0.947 \pm 0.022}$$^\dagger$  & $\mathbf{27.223 \pm 3.353}$$^\dagger$  & $\mathbf{0.051 \pm 0.013}$$^\dagger$         \Tstrut\Bstrut\\
        \hline
        Supervised (Upper Bound) & $\underline{0.996 \pm 0.002}$  & $\underline{39.080 \pm 4.336}$  & $\underline{0.003 \pm 0.002}$    & $\underline{0.954 \pm 0.020}$  & $\underline{27.760 \pm 3.349}$  & $\underline{0.046 \pm 0.013}$         \Tstrut\Bstrut\\
        \hline
        \hline
        \textbf{Evaluation}      & \multicolumn{3}{c}{\textbf{Setting 1 (R=2)}}              &  \multicolumn{3}{c}{\textbf{Setting 2 (R=4)}}                  \Tstrut\Bstrut\\
        \cline{2-7}
        \textbf{T2w}             & SSIM             & PSNR              & NMSE               & SSIM             & PSNR              & NMSE                    \Tstrut\Bstrut\\
        \hline
        Adjoint                  & $0.795 \pm 0.033$  & $15.909 \pm 2.307$  & $0.883 \pm 0.356$    & $0.745 \pm 0.042$  & $14.735 \pm 2.322$  & $1.184 \pm 0.526$         \Tstrut\Bstrut\\
        Gridding/SDC             & $0.866 \pm 0.027$  & $19.170 \pm 2.556$  & $0.402 \pm 0.128$    & $0.800 \pm 0.042$  & $16.712 \pm 2.741$  & $0.743 \pm 0.303$         \Tstrut\Bstrut\\
        CG-SENSE                 & $0.908 \pm 0.019$  & $22.398 \pm 2.454$  & $0.183 \pm 0.035$    & $0.860 \pm 0.032$  & $19.550 \pm 2.833$  & $0.368 \pm 0.121$         \Tstrut\Bstrut\\
        L1-Wavelet               & $0.914 \pm 0.017$  & $22.889 \pm 2.403$  & $0.163 \pm 0.030$    & $0.874 \pm 0.028$  & $20.190 \pm 2.750$  & $0.314 \pm 0.095$         \Tstrut\Bstrut\\
        \hline
        SSDU                     & $0.950 \pm 0.021$  & $26.030 \pm 1.577$  & $0.089 \pm 0.041$    & $0.900 \pm 0.023$  & $23.458 \pm 2.552$  & $0.142 \pm 0.015$         \Tstrut\Bstrut\\
        KDSS                     & $0.980 \pm 0.003$  & $30.490 \pm 2.071$  & $0.028 \pm 0.006$    & $0.944 \pm 0.017$  & $26.513 \pm 2.866$  & $0.071 \pm 0.011$         \Tstrut\Bstrut\\
        DDSS (Ours)              & $\mathbf{0.988 \pm 0.004}$$^\dagger$  & $\mathbf{33.739 \pm 3.324}$$^\dagger$  & $\mathbf{0.013 \pm 0.003}$$^\dagger$    & $\mathbf{0.949 \pm 0.015}$$^\dagger$  & $\mathbf{27.213 \pm 2.900}$$^\dagger$  & $\mathbf{0.060 \pm 0.009}$$^\dagger$         \Tstrut\Bstrut\\
        \hline
        Supervised (Upper Bound) & $\underline{0.996 \pm 0.002}$  & $\underline{39.102 \pm 3.743}$  & $\underline{0.004 \pm 0.002}$    & $\underline{0.953 \pm 0.013}$  & $\underline{27.544 \pm 2.909}$  & $\underline{0.057 \pm 0.011}$         \Tstrut\Bstrut\\
        \hline
    \end{tabular}
\end{table*}

\subsection{Dual-Domain Self-Supervised Learning}
Let $f_{vn}(A, y)$ denote the variational network presented in the previous section, where $A$ is the non-uniform Fourier sampling operator and $y$ is the undersampled k-space measurement. DDSS training and testing pipelines are summarized in Fig. \ref{fig:pipeline}. During training, we randomly partition $y$ into two disjoint sets:
\begin{align}
    y_{p_1} & = S(y, p_1), \\
    y_{p_2} & = S(y, p_2), 
\end{align}
where $S$ is a sampling function with sampling locations $p_1$ and $p_2$. $S$ retrieves the k-space data at location $p_1$ and location $p_2$ to generate the partitioned data $y_{p_1}$ and $y_{p_2}$. $p_1$ and $p_2$ do not share any overlapping coordinates and are randomly generated during training. The partitioned data $y_{p_1}$ and $y_{p_2}$ are then fed into the variational network for parallel reconstruction (with shared weights),
\begin{align}
    x_{p_1} & = f_{vn}(A_{y_{p_1}}, y_{p_1}), \\
    x_{p_2} & = f_{vn}(A_{y_{p_2}}, y_{p_2}). 
\end{align}
In addition to using $y_{p_1}$ and $y_{p_2}$, we also feed the original measurement data before partitioning $y$ into $f_{vn}(\cdot)$ in parallel:
\begin{equation}
    x_{u} = f_{vn}(A, y)
\end{equation}
Our first loss corresponds to a Partition Data Consistency (PDC) loss, which operates in k-space. If the reconstruction network can generate a high-quality image from any undersampled k-space measurements, the k-space data of the images $x_1$, $x_2$, and $x_u$ predicted from the partitions $y_{p_1, p_2}$ and $y$ should all be consistent with the original undersampled k-space data $y$. The predicted k-space data on the original undersampled k-space locations is written as,
\begin{align}
    y_{pred_1} &= Ax_{p_1}, \\
    y_{pred_2} &= Ax_{p_2},  \\
    y_{pred_{u}} &= Ax_{u}. 
\end{align}
Thus, the PDC loss can be formulated as,
\begin{equation}
    \mathcal{L}_{PDC} = ||y_{pred_1} - y||_1 + ||y_{pred_2} - y||_1  + ||y_{pred_u} - y||_1,
\end{equation}
where the first, second, and third terms are the data consistency losses for partitions 1 and 2 and the original undersampled data, respectively.

Our second loss is an Appearance Consistency (AC) loss which operates in the image domain. First, we regularize the reconstructions from the partitions $y_{p_1}$ and $y_{p_2}$ to be consistent with each other at the image-level. Second, we assume that the reconstruction from $y$ should also be consistent with the reconstructions of $y_{p_1}$ and $y_{p_2}$. To enforce that, the AC loss is computed on both image intensities and image gradients for improved anatomical clarity,
\begin{equation}
    \mathcal{L}_{AC} = \lambda_{img} \mathcal{L}_{img} + \lambda_{grad} \mathcal{L}_{grad},
\end{equation}
where,
\begin{equation}
    \mathcal{L}_{img} = ||x - x_{p_1}||_1 + ||x - x_{p_2}||_1  + ||x_{p_1} - x_{p_2}||_1 \ 
\end{equation}
\begin{align}
    \mathcal{L}_{grad} &  = ||\nabla_v x_u - \nabla_v x_{p_1}||_1 + ||\nabla_h x_u - \nabla_h x_{p_1}||_1, \\
             & + ||\nabla_v x_u - \nabla_v x_{p_2}||_1 + ||\nabla_h x_u - \nabla_h x_{p_2}||_1, \nonumber \\
             & + ||\nabla_v x_{p_1} - \nabla_v x_{p_2}||_1 + ||\nabla_h x_{p_1} - \nabla_h x_{p_2}||_1, \nonumber
\end{align}
where $\nabla_v$ and $\nabla_h$ are spatial intensity gradient operators in x and y directions, respectively. We set $\lambda_{img} = 2$ and $\lambda_{grad} = 1$. Combining the PDC loss in k-space and the AC loss in the image domain, our total loss is,
\begin{equation} \label{eq:loss_ddss}
    \mathcal{L}_{tot} = \mathcal{L}_{AC} + \lambda_{PDC} \mathcal{L}_{PDC}
\end{equation}
where $\lambda_{PDC} = 10$ is used to balance the scale between k-space and image domain losses, which is selected by hyper-parameter search (Fig. \ref{fig:plot_weight}).

\section{Experiments}
\subsection{Data Preparation}
We evaluated the proposed method on both simulated and real non-Cartesian data. For the simulation studies, we randomly select 505 T1-weighted and 125 T2-weighted 3D brain MR images from the Human Connectome Project (HCP) \citep{van2013wu} with no subject overlap. The volumes were first resampled to $1.5 \times 1.5 \times 5 mm^3$ to match common clinical resolutions. We consider a 2D non-Cartesian multi-coil data acquisition protocol, where 8 coil sensitivity maps are analytically generated. We note that the publicly-available HCP dataset is magnitude-only and does not come with phase information. To this end, we add realistic phases to the magnitude images to create fully complex-valued target images for the experiments, which generates complex-valued multi-coil 3D images for the simulation study.

To generate non-Cartesian undersampled data, we use a non-Cartesian variable density sampling pattern where the sampling density decays from the k-space center at a quadratic rate \citep{lustig2008compressed} with the central 10\% of the k-space oversampled at $1.25$ times the Nyquist rate. We generate two sampling trajectory settings with target acceleration factor $R \in \{2, 4\}$. 476 T1-weighted and 104 T2-weighted images are used for training and 29 T1-weighted and 21 T2-weighted MR images are used for evaluation.

For real-world low-field MRI studies, we collect 119 FLAIR and 125 FSE-T2w 3D brain MR images acquired using a Hyperfine Swoop\textsuperscript{TM} portable MRI system\footnote{https://www.hyperfine.io/} with a field strength of 64 mT. An in-house 1 transmitter / 8 receivers phased array coil is implemented in the system. Walsh’s method \citep{walsh2000adaptive} was used to estimate the coil sensitivity maps as required for the reconstruction. Due to the low field strength and real-world deployment, these images are more difficult to reconstruct due to higher levels of noise and imaging system artifacts. Both FLAIR and FSE-T2w images of resolution $1.6 \times 1.6 \times 5 mm^3$ are acquired using a non-Cartesian variable density sampling pattern with an acceleration factor of 2 and 4 and are reconstructed slice-wise. 106 FLAIR and 112 FSE-T2w images were used for training and 13 FLAIR and 13 FSE-T2w MR images are used for evaluation.

\subsection{Implementation Details}
We implement all methods in Tensorflow and perform experiments using an NVIDIA Tesla M60 GPU with 8GB memory. The Adam solver \citep{kingma2014adam} is used to optimize our models with $lr = 3 \times 10^{-5}$, $\beta_{1} = 0.9$, and $\beta_{2} = 0.999$. We use a batch size of 8 and train all models for 200 epochs. To trade off training time versus reconstruction quality, the default number of iterations in the non-Cartesian reconstruction network was set to 6 unless otherwise noted. We use the adjoint for $f_{init}$. We initialize the forward and adjoint operator based on \citet{fessler2003nonuniform} with an oversampling factor of 2. During training, the undersampled data partitioning rate is randomly generated between $[0.2, 0.8]$ which was empirically determined to provide better results during initial prototyping. 

\subsection{Baselines and Evaluation Strategies}
\textbf{Simulated HCP.} For the simulated non-Cartesian HCP data with ground truth reconstruction available, we quantify reconstructed image quality via the structural similarity index (SSIM), peak signal-to-noise ratio (PSNR), and normalized mean squared error (NMSE). Benchmarks are run against previous methods including SSDU \citep{yaman2020self}, Compressed Sensing with $L_1$-Wavelet sparsity \citep{gu2021compressed}, CG-SENSE \citep{maier2021cg}, Gridding \citep{zwart2012efficient}, and adjoint-only reconstruction. For fair comparison, the same non-Cartesian reconstruction network was deployed in SSDU. As an upper bound, we also compared our DDSS against a supervised strategy where the same reconstruction network was trained in a fully supervised fashion with ground truth available. As ablations, we further evaluate the advantages of dual-domain training, where we compared against a k-space domain only self-supervised (KDSS) model where only the PDC loss is used for training. \\

\textbf{Real-world low-field MRI.}  For the acquired low-field non-Cartesian data with no ground truth available, image quality assessment is performed via a reader study including three experienced radiologists. The radiologists are asked to compare DDSS to CG-SENSE \citep{maier2021cg}, a reconstruction algorithm implemented by the Hyperfine system which robustly offers high quality reconstruction. During the reader study, pairs of CG-SENSE and DDSS reconstruction with an effective acceleration rate of 2 were shown, where the readers are blinded to the reconstruction method. The evaluations are performed using a five-point Likert scale to rate the image quality and the consistency in diagnosis outcomes. The image quality is evaluated in terms of noise, sharpness, and overall perceptual quality between DDSS and CG-SENSE, with a rating scale of 2 being far better, 1 being better, 0 being the same, -1 being worse, -2 being far worse. The consistency in diagnosis outcomes is evaluated based on the agreement on giving consistent diagnoses between the two methods' reconstructions in terms of contrast, geometric fidelity, and presence of artifacts (2: strongly agree, 1: agree, 0: neutral, -1: disagree, -2: strongly disagree). Specifically, for contrast, the reader is asked to judge whether the contrast between different anatomical features in both image reconstructions is the same. For geometric fidelity, the reader assesses whether whether CG-SENSE and DDSS agree on how fine details in underlying anatomy are rendered. For artifact assessment, the reader is evaluates whether CG-SENSE and DDSS agree on the presence or absence of imaging artifacts, such as motion and ghosting.

\subsection{Results on Simulated Non-Cartesian Data}
For the simulation studies, quantitative results are summarized in Table \ref{tab:comp_methods}. With acceleration $R=2$ on T1w reconstruction, DDSS achieves SSIM$=0.988$ and PSNR=$33.678$ which are significantly higher than all previous baseline methods, including SSDU \citep{yaman2020self} with SSIM=$0.960$ and PSNR=$26.301$. When comparing the ablation KDSS against DDSS, we find that the AC loss further boosts reconstruction performance from $30.454$ to $33.678$ in terms of PSNR and from $0.981$ to $0.988$ in terms of SSIM, thus narrowing the gap in performance to the fully supervised upper bound with PSNR=$39.090$ and SSIM=$0.996$. Similar trends are observed for the T2w reconstruction experiments where DDSS outperforms previous unsupervised (conventional or deep learning-based) baseline methods, achieving a PSNR of $33.739$ and SSIM of $0.988$. While overall performance decreases for all methods as the acceleration factor increases from $R=2$ to $R=4$, DDSS still outperforms previous methods on both T1w and T2w reconstruction tasks in terms of PSNR, SSIM and NMSE. Interestingly, for $R=2$, PSNR standard deviation marginally increases with increased mean PSNR performance, but we do not find similar patterns in the $R=4$ experiments. Since DDSS, KDSS, and SSDU are implemented with the same reconstruction network, they share the same reconstruction time that is $<0.01s$, and the number of parameters ($30M$).

Qualitative comparisons of T1w reconstructions and T2w reconstructions under R=2 and R=4 are visualized in Fig. \ref{fig:compare_simulation_T1_methods} and Fig. \ref{fig:compare_simulation_T2_methods}, respectively. We find that conventional methods such as CG-SENSE and $L_1$-Wavelet can significantly reduce aliasing as compared to the gridding reconstructions. SSDU further reduces the artifacts with decreased residual reconstruction errors as compared to conventional methods. On the other hand, DDSS reconstruction resulted in the lowest overall errors for both T1w and T2w reconstructions across all previous conventional and self-supervised methods. Even though the supervised model using fully sampled ground truth during training achieves the best quantitative results, the reconstructions from the supervised model and DDSS are comparable qualitatively.

\begin{figure}[tb!]
\centering
\includegraphics[width=0.483\textwidth]{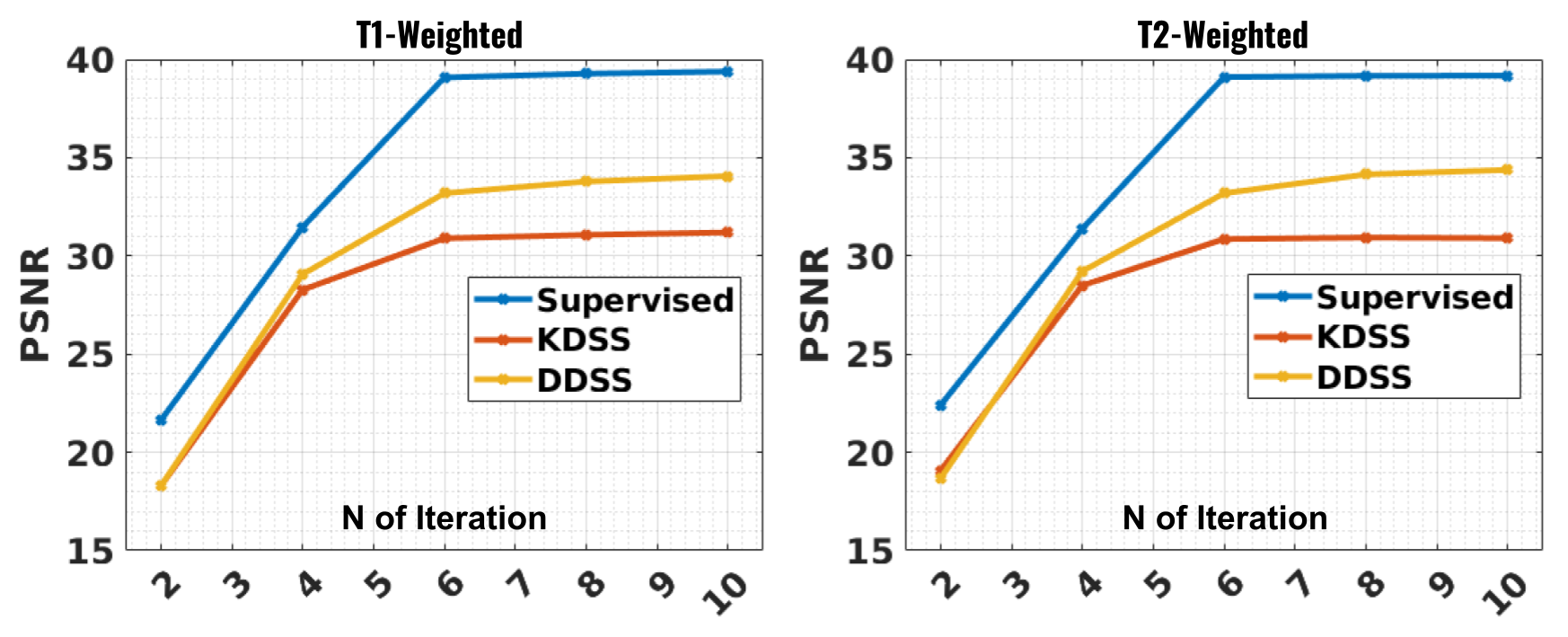}
\caption{The impact of the number of iterations in the non-Cartesian reconstruction network for DDSS reconstruction on the Simulated HCP dataset.}
\label{fig:plot_unroll}
\end{figure}

We then study the impact of the number of iterations in the non-Cartesian reconstruction network (Fig. \ref{fig:plot_unroll}). We find that DDSS consistently outperforms KDSS at different settings of the number of iterations and achieved closer performance to the fully supervised model. For the DDSS, the reconstruction performance starts to asymptotically converge after $6$ iterations. Experiments on T1w and T2w reconstructions show similar behavior when varying the number of iteration from 2 to 10.

\begin{figure}[htb!]
\centering
\includegraphics[width=0.483\textwidth]{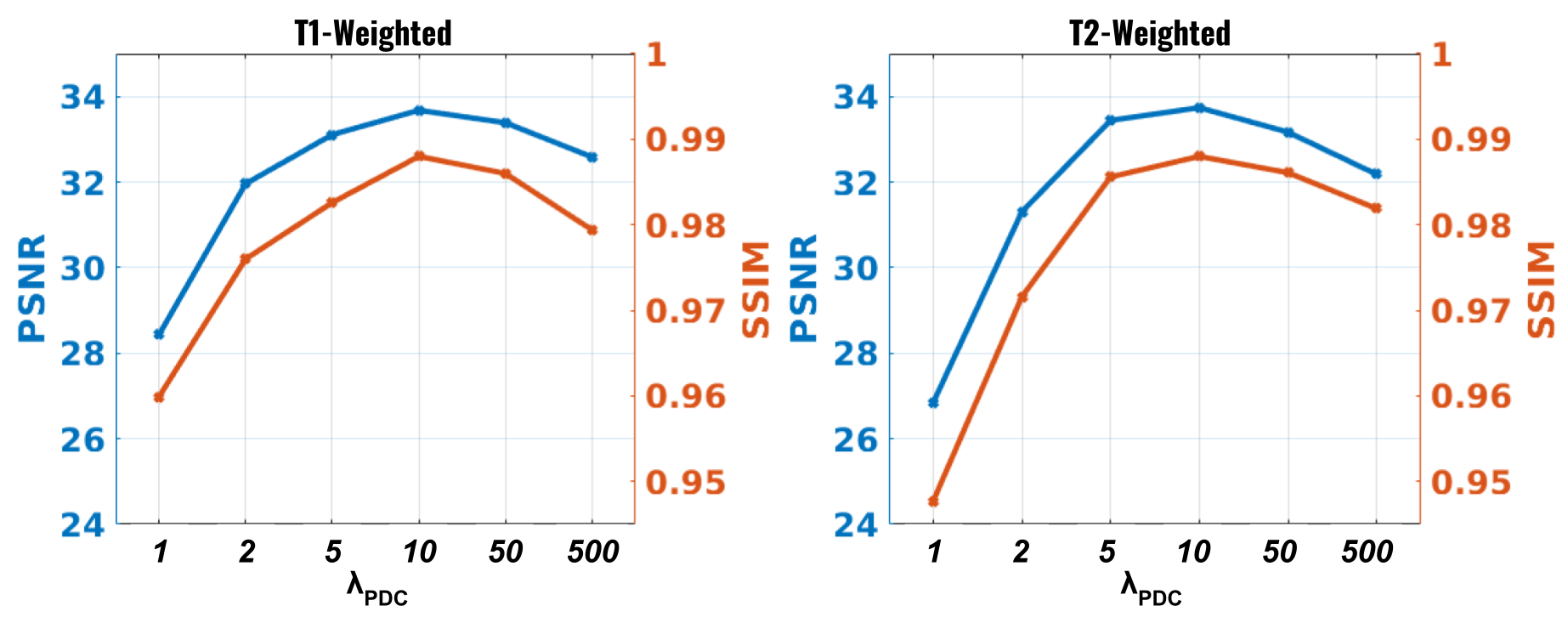}
\caption{The impact of $\lambda_{PDC}$ used in DDSS training (Eq. \ref{eq:loss_ddss}). Higher is better.}
\label{fig:plot_weight}
\end{figure}

We further investigate the impact of $\lambda_{PDC}$ used in DDSS training (Fig. \ref{fig:plot_weight}). When sweeping $\lambda_{PDC}$ from 1 to 500, the T1w and T2w reconstruction performance of DDSS is optimized when $\lambda_{PDC}$ is set to 10, which is used as the default hyper-parameter in our experiments. Notably, using $\lambda_{PDC} = 0$, i.e. only using the AC loss in the loss function, was unable to converge due to training instability indicating that both components of the overall DDSS loss are synergistic.

\begin{figure*}[htb!]
\centering
\includegraphics[width=0.8\textwidth]{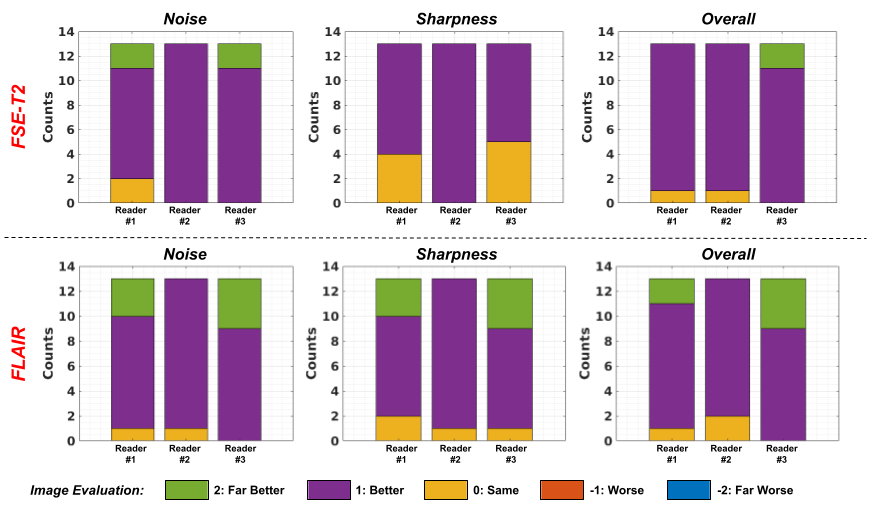}
\caption{The image-quality expert assessments from the clinical radiologist reader study on the real low-field MRI dataset. Image-quality is evaluated in terms of noise, sharpness, and overall quality using a 5-point ordinal scale, comparing DDSS to CG-SENSE \citep{maier2021cg} reconstructions.}
\label{fig:reader_study_NSO}
\end{figure*}

\begin{figure*}[htb!]
\centering
\includegraphics[width=0.8\textwidth]{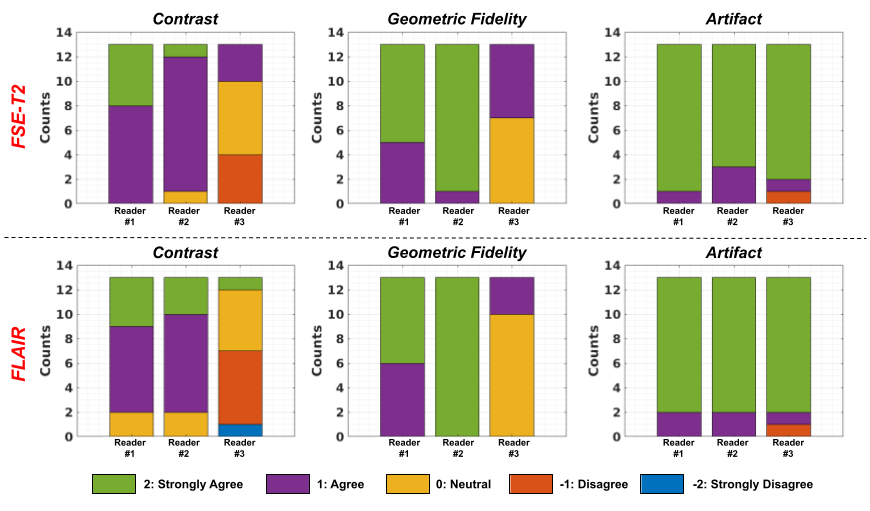}
\caption{The diagnostic-consistency expert assessments from the clinical radiologist reader study on the real low-field MRI dataset. Diagnosis consistency is evaluated in terms of contrast, geometric fidelity, and presence/absence of artifacts using a 5-point ordinal scale, comparing DDSS to CG-SENSE \citep{maier2021cg} reconstructions.}
\label{fig:reader_study_CGA}
\end{figure*}

\begin{figure*}[htb!]
\centering
\includegraphics[width=1.000\textwidth]{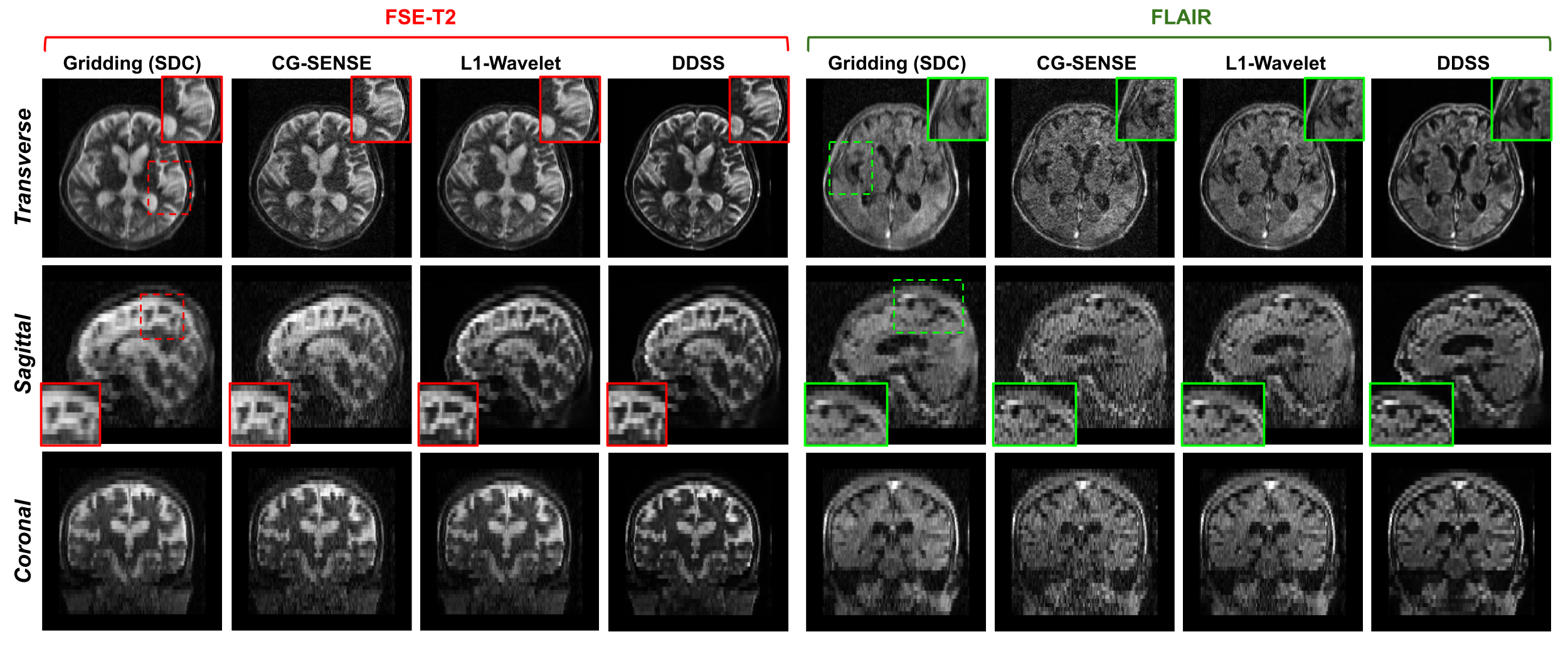}
\caption{Qualitative comparisons of FSE-T2w and FLAIR reconstructions from the real data acquired from a low-field (64mT) MRI system. The subject was diagnosed with a hemorrhagic stroke with lacunar infarcts. The DDSS reconstruction is compared to Gridding, L1-Wavelet, and CG-SENSE \citep{maier2021cg} reconstructions of this subject.}
\label{fig:real_data_case1}
\end{figure*}

\begin{figure*}[htb!]
\centering
\includegraphics[width=1.000\textwidth]{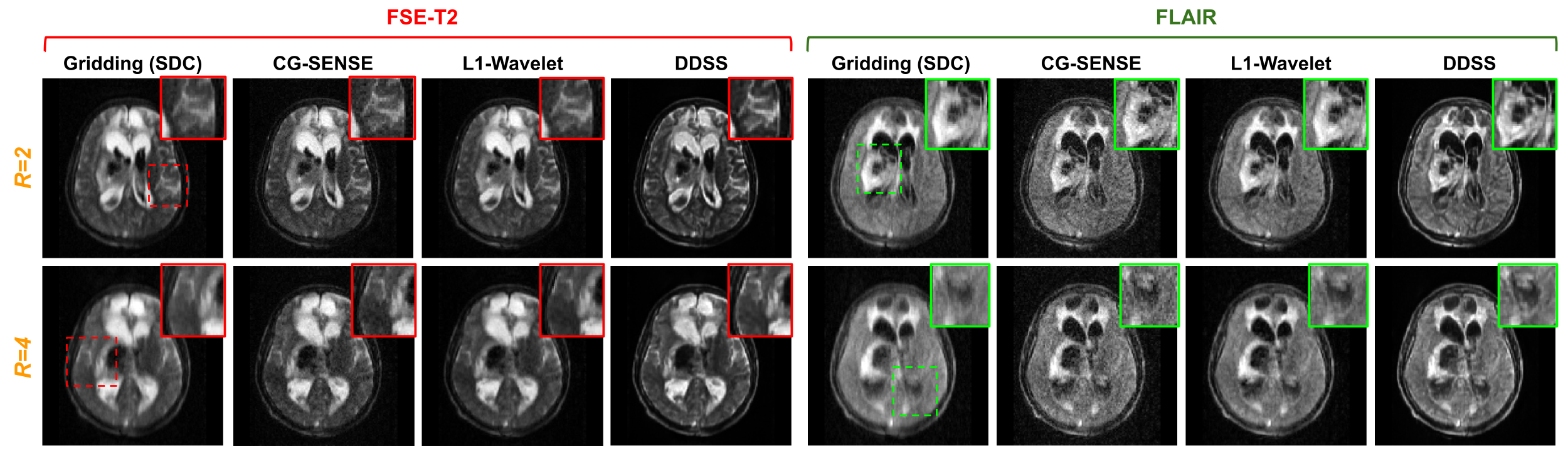}
\caption{Qualitative comparisons of FSE-T2w and FLAIR reconstructions from the real data acquired from a low-field (64mT) MRI system. The subject was diagnosed with a hemorrhagic stroke with associated midline shift. The DDSS reconstruction is compared to Gridding, L1-Wavelet, and CG-SENSE \citep{maier2021cg} reconstructions of this subject under both R=2 and R=4 acceleration factors.}
\label{fig:real_data_case2}
\end{figure*}

\subsection{Results on Real Data}
For the real-world dataset, the data acquired from the low-field MRI scanner is non-Cartesian and accelerated with no ground truth available \citep{Mazurek2021}. We therefore evaluated DDSS performance on this dataset using expert studies. Specifically, we trained DDSS on the real data and compared it against the default CG-SENSE reconstruction \citep{maier2021cg} from the system. The reader study for the image quality assessment is summarized in Fig. \ref{fig:reader_study_NSO} and Tbl. \ref{tab:reader_study_NSO}.

All 3 radiologists rated the DDSS FSE-T2 and FLAIR reconstructions to be better than or the same as the reconstructions from the CG-SENSE, in terms of noise, sharpness, and overall image quality. For noise, sharpness, and overall image quality, the averaged rating scores of the 3 radiologists were $1.05$, $0.76$, and $1.00$, respectively for FSE-T2 and $1.12$, $1.07$, and $1.07$ respectively for FLAIR reconstruction.

\begin{table} [htb!]
\footnotesize
\centering
\caption{The averaged scores of the 3 radiologists from the image-quality reader study on both FSE-T2 and FLAIR reconstructions. Higher is better.}
\label{tab:reader_study_NSO}
    \begin{tabular}{l|c c c|c}
        \hline
        \textbf{FSE-T2}        & \textbf{Reader \#1}      & \textbf{Reader \#2}      & \textbf{Reader \#3}      & \textbf{Average}        \Tstrut\Bstrut\\
        \hline
        \textbf{Noise}         & $1.00 \pm 0.57$  & $1.00 \pm 0.00$  & $1.15 \pm 0.37$  & $1.05 \pm 0.39$  \Tstrut\Bstrut\\
        \hline
        \textbf{Sharpness}     & $0.69 \pm 0.48$  & $1.00 \pm 0.00$  & $0.61 \pm 0.50$  & $0.76 \pm 0.42$   \Tstrut\Bstrut\\
        \hline
        \textbf{Overall}       & $0.92 \pm 0.27$  & $0.92 \pm 0.27$  & $1.15 \pm 0.37$  & $1.00 \pm 0.32$   \Tstrut\Bstrut\\
        \hline
        \hline
        \textbf{FLAIR}         & \textbf{Reader \#1}      & \textbf{Reader \#2}      & \textbf{Reader \#3}      & \textbf{Average}        \Tstrut\Bstrut\\
        \hline
        \textbf{Noise}         & $1.15 \pm 0.55$  & $0.92 \pm 0.27$  & $1.30 \pm 0.48$  & $1.12 \pm 0.46$  \Tstrut\Bstrut\\
        \hline
        \textbf{Sharpness}     & $1.07 \pm 0.64$  & $0.92 \pm 0.27$  & $1.23 \pm 0.59$  & $1.07 \pm 0.53$   \Tstrut\Bstrut\\
        \hline
        \textbf{Overall}       & $1.07 \pm 0.49$  & $0.84 \pm 0.37$  & $1.30 \pm 0.48$  & $1.07 \pm 0.48$   \Tstrut\Bstrut\\
        \hline
    \end{tabular}
\end{table}

The reader study for the consistency in diagnoses is summarized in Fig. \ref{fig:reader_study_CGA} and Tbl. \ref{tab:reader_study_CGA}. The first and second radiologists agreed that consistent diagnoses were attainable from the DDSS and CG-SENSE reconstructions for the majority of the study, in terms of contrast, geometric fidelity, and the presence of artifacts. Notably, the third radiologist rated neutral or disagree in terms of contrast consistency between the two methods and rated agree or neutral in terms of geometric consistency for majority of the cases.
In the diagnosis consistency study, the averaged rating scores of the 3 radiologists for contrast, geometric fidelity, and presence of artifacts are $0.76$, $1.33$, and $1.79$, respectively, for FSE-T2 and $0.58$, $1.25$, and $1.79$, respectively, for FLAIR reconstruction.

\begin{table} [htb!]
\footnotesize
\centering
\caption{The averaged scores of the 3 radiologists from the diagnosis consistency reader study on both FSE-T2 and FLAIR reconstructions. Higher is better.}
\label{tab:reader_study_CGA}
    \begin{tabular}{l|c c c|c}
        \hline
        \textbf{FSE-T2}        & \textbf{Reader \#1}      & \textbf{Reader \#2}      & \textbf{Reader \#3}      & \textbf{Average}        \Tstrut\Bstrut\\
        \hline
        \textbf{Contrast}      & $1.38 \pm 0.50$  & $1.00 \pm 0.40$  & $-0.07 \pm 0.75$  & $0.76 \pm 0.84$  \Tstrut\Bstrut\\
        \hline
        \textbf{Geometry}      & $1.61 \pm 0.50$  & $1.92 \pm 0.27$  & $0.46 \pm 0.51$  & $1.33 \pm 0.77$   \Tstrut\Bstrut\\
        \hline
        \textbf{Artifact}      & $1.92 \pm 0.27$  & $1.76 \pm 0.43$  & $1.69 \pm 0.85$  & $1.79 \pm 0.57$   \Tstrut\Bstrut\\
        \hline
        \hline
        \textbf{FLAIR}         & \textbf{Reader \#1}      & \textbf{Reader \#2}      & \textbf{Reader \#3}      & \textbf{Average}        \Tstrut\Bstrut\\
        \hline
        \textbf{Contrast}      & $1.15 \pm 0.68$  & $1.07 \pm 0.64$  & $-0.46 \pm 0.96$  & $0.58 \pm 1.06$  \Tstrut\Bstrut\\
        \hline
        \textbf{Geometry}      & $1.53 \pm 0.51$  & $2.00 \pm 0.00$  & $0.23 \pm 0.43$  & $1.25 \pm 0.84$   \Tstrut\Bstrut\\
        \hline
        \textbf{Artifact}      & $1.84 \pm 0.37$  & $1.84 \pm 0.37$  & $1.69 \pm 0.85$  & $1.79 \pm 0.57$   \Tstrut\Bstrut\\
        \hline
    \end{tabular}
\end{table}

Qualitative results are presented in Fig. \ref{fig:real_data_case1} and Fig. \ref{fig:real_data_case2}, where we visualize results from FSE-T2w and FLAIR scans of two subjects presenting with hemorrhagic strokes. In addition to visualizing the CG-SENSE reconstruction, we also visualize the Gridding and L1-Wavelet reconstruction for qualitative comparisons. As we can see from Fig. \ref{fig:real_data_case1} with R=2, the Gridding reconstructions suffer from blurring due to the accelerated data acquisition protocols. While the L1-Wavelet and CG-SENSE methods can reduce blurring, the proposed self-supervised DDSS reconstructions produce much sharper image quality leading to enhanced visualization of neuroanatomy. Similar observation are made in Fig. \ref{fig:real_data_case2}, where DDSS provides reconstructions with better contrast, sharpness, and lower noise under both R=2 and R=4 acceleration factors.

\section{Discussion} 

In this work, we developed a novel dual-domain self-supervised (DDSS) approach for accelerated non-Cartesian MRI reconstruction. Specifically, we proposed to train a non-Cartesian MRI reconstruction network in both image and k-space domains in a self-supervised fashion. We overcame two major difficulties in MRI reconstruction. First, the proposed self-supervised method allows the MRI reconstruction network to be trained without using any fully sampled MRI data, instead of relying on large-scale under/fully-sampled paired data for reconstruction network training, which is infeasible if the MRI system has accelerated acquisition protocols. Second, DDSS is applicable to non-Cartesian MRI reconstruction, a relatively understudied problem for deep reconstruction networks.

\subsection{Experimental Summaries}
We first demonstrate that DDSS can reconstruct high-quality images on the simulated non-Cartesian MRI dataset (Tab \ref{tab:comp_methods}). First of all, the DDSS can achieve significantly better reconstruction performance than baseline reconstruction methods, including previous conventional and self-supervised methods. Secondly, we found combining image domain self-supervision with k-space self-supervision can significantly boost the reconstruction performance, implying the synergy of dual-domain self-supervision enhances the reconstruction learning of DDSS. Qualitatively, we can observe from Fig \ref{fig:compare_simulation_T1_methods} and Fig. \ref{fig:compare_simulation_T2_methods} that the important anatomy structures are visually consistent with the ground truth. Ablation study on the impact of $\lambda_{PDC}$ which controls the balance between the k-space self-supervision loss and the image-domain self-supervision loss shows that it is important to properly select this hyper-parameter to achieve optimal performance. We also found that setting $\lambda_{PDC}=0$, i.e. using only image-domain self-supervision during the training, cannot properly converge the loss, thus the performance under this setting is not reported. Finally, we also demonstrate successful applications on a real dataset, where the low-field non-Cartesian MRI data is acquired using the Hyperfine Swoop system with only undersampled non-Cartesian data available (Figs. \ref{fig:real_data_case1} and \ref{fig:real_data_case2}). Our reader studies on image-quality (Fig. \ref{fig:reader_study_NSO} and Tab. \ref{tab:reader_study_NSO}) and diagnosis consistency (Fig. \ref{fig:reader_study_CGA} and Tbl. \ref{fig:reader_study_NSO}) show that the DDSS can provide superior image quality and highly consistent diagnosis results as compared to the conventional method deployed in the system, i.e. CG-SENSE.

\subsection{Limitations and Future Work} 
The presented work has opportunities for improvements that are the subject of our ongoing work. 
\begin{itemize}
\item While DDSS performance does not currently exceed the performance of full supervision, several future modifications could potentially further increase its performance. First, the current non-Cartesian reconstruction network is based on gradient descent where the data consistency constraint could be further enforced. Using a conjugate gradient-based architecture, such as \citet{aggarwal2018modl}, could potentially further improve DDSS performance. Similarly, deploying a density-compensated primal dual network \citep{ramzi2022nc} with a density-compensated data consistency operation in DDSS could also improve reconstruction.
\item DDSS only imposes image and k-space self-supervised losses on the final output of the non-Cartesian reconstruction network, whereas deep supervision on each cascade output \citep{zhou2020dudornet} could be implemented to potentially further improve its performance. 
\item Even though this work use a UNet \citep{ronneberger2015u} as the backbone network in the non-Cartesian reconstruction network, we do not claim that this is an optimal backbone for reconstruction. As the DDSS backbone network is interchangeable, other state-of-the-art image restoration networks, such as OUCNet \citep{guo2021over}, ResViT \citep{dalmaz2021resvit}, and DCCT \citep{zhou2022dsformer}, could use the DDSS loss functions which could lead to improved reconstruction. Deploying dual-domain reconstruction networks, such as DuDoRNet \citep{zhou2020dudornet} and MDReconNet \citep{ran2020md}, in DDSS and extending the current 2D framework to 3D, could also potentially improve the reconstruction performance. 
\item While the coil sensitivity map is assumed to be known or straightforward to estimate in this work, previous work \citep{sriram2020end} has attempted to use another sub-network for coil sensitivity map estimation. Integrating deep learning-based coil sensitivity map prediction into DDSS will be investigated in future work. 
\item This work focused on the non-Cartesian variable density sampling pattern in order to facilitate experimentation with the real clinical images generated by Hyperfine system. Investigating more diverse non-Cartesian sampling patterns, such as spiral interleaves and radial spokes \citep{tsao2012mri} and their compatibility with DDSS, will be included in future work.
\item The experiments in this paper focus on clinical scenarios, in which extremely high fidelity reconstruction is required, and as such the acceleration factors were investigated at ($2\times$ and $4\times$). On the other hand, in research scenarios, much higher acceleration factors are often considered in order to probe the limit of the methods. From Table \ref{tab:comp_methods}, we see that, while statistically significant, the performance gains are not as strong for $R=4$ as they are for $R=2$. It is plausible that even higher acceleration rates may further reduce the impact of self-supervision via k-space redundancy, especially in comparison to a fully-supervised model which still utilizes the fully-sampled ground truth for supervision. Therefore, future work will investigate the sensitivity of the DDSS framework to higher acceleration factors and potential improvements.
\end{itemize}

\section{Conclusion} 
This paper presented a dual-domain self-supervised learning method for training a non-Cartesian deep MRI reconstruction model \textit{without} using any fully sampled data. Novel loss functions leveraging self-supervision in both the k-space and image domains were developed leading to improved reconstructions. Experimental results on a simulated accelerated non-Cartesian dataset demonstrated that DDSS can generate highly accurate reconstructions that approach the fidelity of the fully supervised reconstruction. Finally, the proposed framework was shown to successfully scale to the reconstruction of challenging real MRI data from a portable low-field 0.064T MRI scanner, where fully sampled data is unavailable. These DDSS improvements were assessed by expert radiologists in a user study measuring image quality and diagnostic consistency and were found to outperform traditional reconstruction methods.

\section*{Acknowledgments}
The authors thank Dr. Ardavan Saeedi for valuable discussions and suggestions and the radiologists for participating in our expert study. 

\section*{Declaration of Competing Interest}
The authors declare that they have no known competing financial interests or personal relationships that could have appeared to influence the work reported in this paper.

\section*{Credit authorship contribution statement }
\textbf{Bo Zhou}: Conceptualization, Methodology, Software, Visualization, Validation, Formal analysis, Writing original draft.
\textbf{Jo Schlemper}: Conceptualization, Methodology, Software, Visualization, Validation, Formal analysis, Writing - review and editing, Supervision.
\textbf{Neel Dey}: Conceptualization, Writing - review and editing.
\textbf{Seged Sadegh Mohseni Salehi}: Conceptualization, Methodology, Software, Writing - review and editing.
\textbf{Kevin Sheth}: Data preparation, Writing - review and editing.
\textbf{Chi Liu}: Writing - review and editing.
\textbf{James S. Duncan}: Writing - review and editing.
\textbf{Michal Sofka}: Conceptualization, Methodology, Software, Visualization, Validation, Formal analysis, Writing - review and editing, Supervision.

\bibliographystyle{model2-names.bst}\biboptions{authoryear}
\bibliography{refs}

\end{document}